\documentclass[12pt,a4paper,final]{iopart}

\usepackage{iopams}  
\usepackage{graphicx}
\usepackage[breaklinks=true,colorlinks=true,linkcolor=blue,urlcolor=blue,citecolor=blue]{hyperref}
\usepackage{breqn}
\usepackage{subfigmat}
\usepackage{xcolor}
\usepackage{setspace}

\expandafter\let\csname equation*\endcsname\relax
\expandafter\let\csname endequation*\endcsname\relax

\begin{document}

\begin{spacing}{2}

\title{Validation of GBS plasma turbulence simulation of the TJ-K stellarator}

\author{A. J. Coelho, J. Loizu, P. Ricci}
\address{Ecole Polytechnique Fédérale de Lausanne (EPFL), Swiss Plasma Center (SPC), CH-1015 Lausanne, Switzerland}
\author{M. Ramisch$^1$, A. Köhn-Seemann$^1$, G. Birkenmeier$^{2,3}$, K. Rahbarnia$^4$}
\address{1. Institute of Interfacial Process Engineering and Plasma Technology IGVP, University of Stuttgart, 70569 Stuttgart, Germany\\
2. Max Planck Institute for Plasma Physics, Boltzmannstr. 2, 85748 Garching, Germany\\
3. Physik-Department E28, Technische Universität München, James-Franck-Str. 1, 85748 Garching, Germany\\
4. Max Planck Institute for Plasma Physics, 17491 Greifswald, Germany}

\ead{antonio.coelho@epfl.ch}

\begin{abstract}
We present a validation of a three-dimensional, two-fluid simulation of plasma turbulence in the TJ-K stellarator, a low temperature plasma experiment ideally suited for turbulence measurements. The simulation is carried out by the GBS code, recently adapted to simulate 3D magnetic fields. The comparison shows that GBS retrieves the main turbulence properties observed in the device, namely the fact that transport is dominated by fluctuations with low poloidal mode number. The poloidal dependence of the radial $\text{E}\times\text{B}$ turbulent flux is compared on a poloidal plane with elliptical flux surfaces, where a very good agreement between experiment and simulation is observed, and on another with triangular flux surfaces, which shows a poorer comparison. The fluctuation levels in both cases are underestimated in the simulations. The equilibrium density profile is well retrieved by the simulation, while the electron temperature and the electrostatic potential profiles, which are very sensitive to the strength and localization of the sources, do not agree well with the experimental measurements. 
\end{abstract}

\newpage

\section{Introduction}

As stellarators are becoming a viable option for a fusion reactor~\cite{TS_Pederson_experimental_confirmation_W7X,W7X_neoclassical_nature}, fluid codes are being extended to non-axisymmetric magnetic field geometries to study the properties of plasma turbulence in the stellarator boundary. BSTING, an extension of the BOUT++ code~\cite{FCI_BOUT}, simulated seeded filaments in a rotating ellipse~\cite{BSTING}. More recently, the first global flux-driven simulations of a stellarator, performed by using the GBS code~\cite{paolo_GBS,GBS_cite_Maurizio,halpern}, considered a vacuum magnetic field generated with the Dommaschk potentials, reporting important differences with respect to tokamak simulations, namely the existence of a low-$m$ mode, where $m$ is the poloidal mode number, dominating the turbulent transport~\cite{gbs_stellarators}. Such surprising result calls for the validation of turbulence simulations in stellarators.

In this paper, we present the first validation of a simulation of plasma turbulence in a stellarator configuration against experimental measurements. We compare helium discharges carried out in the TJ-K stellarator with a simulation performed using the GBS code. TJ-K is a stellarator experiment ideally suited for a detailed comparison with simulations~\cite{TJ-K}. Because of the low plasma density and electron temperature, Langmuir probes can access the entire plasma volume and provide equilibrium as well as turbulence measurements that can be easily compared with simulations. In addition, since collisionality in TJ-K is large in the whole plasma volume, the fluid equations evolved by GBS are valid both in the core and in the boundary regions. Finally, the small size of the machine makes its simulation attractive from the point of view of the computational cost. 

Previous turbulence modelling of TJ-K was accomplished either through modified Hasegawa-Wakatani models in slab geometry~\cite{TJK_Garland_1,TJK_Garland_2} or by using a fluid model in a simplified geometry with the characteristic plasma parameters of TJ-K~\cite{TJK_DALF3}. Due to their simplicity, a detailed one-to-one comparison of the simulations against experimental results was not attempted. The present work leverages previous validations of GBS against experiments carried out in axisymmetric configurations~\cite{torpex_validation1,torpex_validation2,torpex_galassi,diego_X21}. Thanks to the full-f nature of the GBS simulation code, which does not make a separation between background and fluctuations, we validate equilibrium as well as fluctuating quantities (density and electrostatic potential).

The paper is organized as follows. Section 2 describes the TJ-K experiment. In Section 3, the physical model implemented in GBS is presented. In Section 4, the simulation results are presented and their validation with the TJ-K experiment is reported. Finally, we discuss our results and draw our conclusions in Section 5.

\section{The TJ-K experiment}

\begin{figure}[]
 \centering
 \includegraphics[width=16cm]{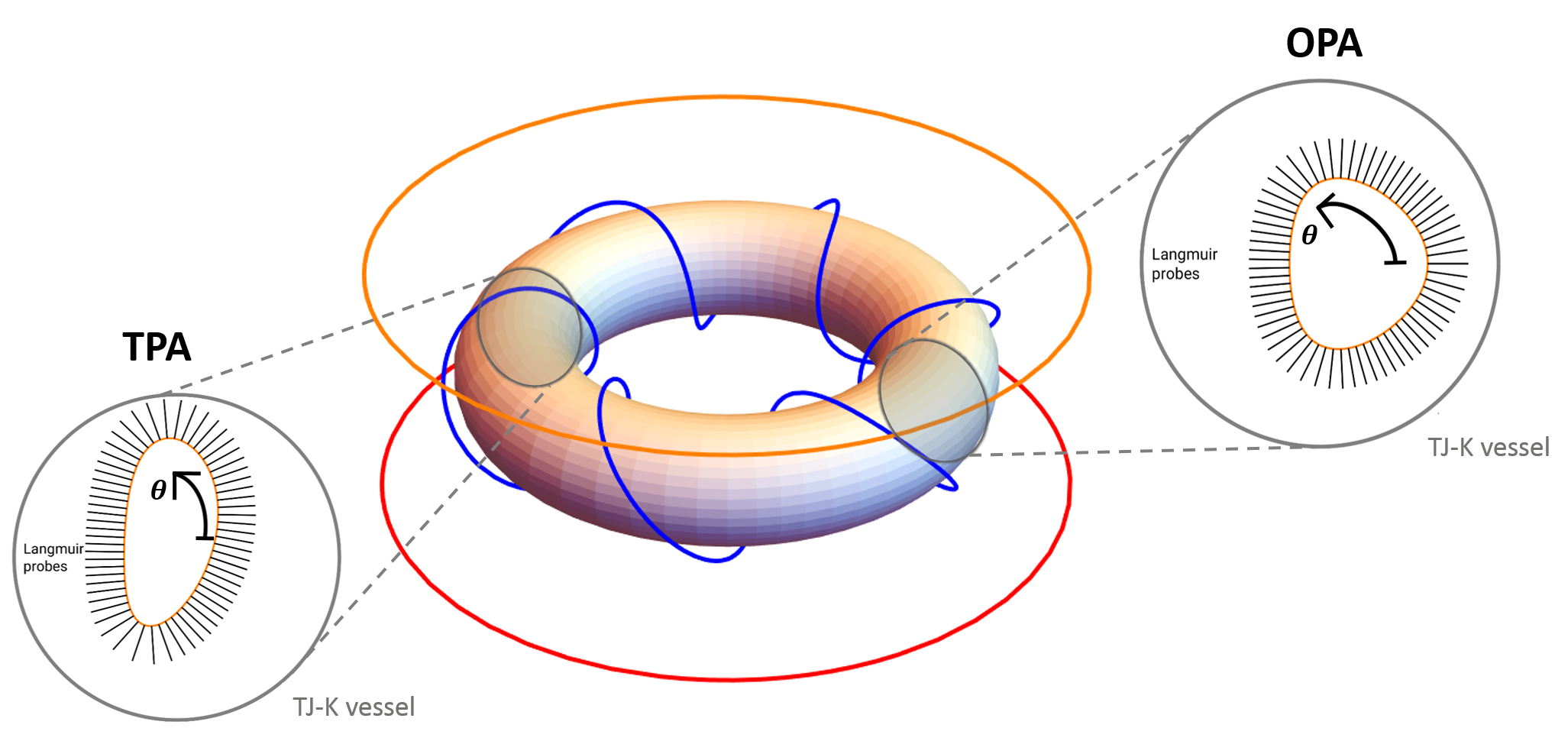}
 \caption{Schematics of the TJ-K experiment. The magnetic field is generated by a helical coil (blue) and two vertical-field circular coils (orange and red). Two multi-Langmuir probe arrays are distributed poloidally along the same flux surface at two different toroidal angles (TPA and OPA).}
 \label{fig:TJK_esquema}
\end{figure}
\begin{figure}[]
 \centering
 \includegraphics[width=16cm]{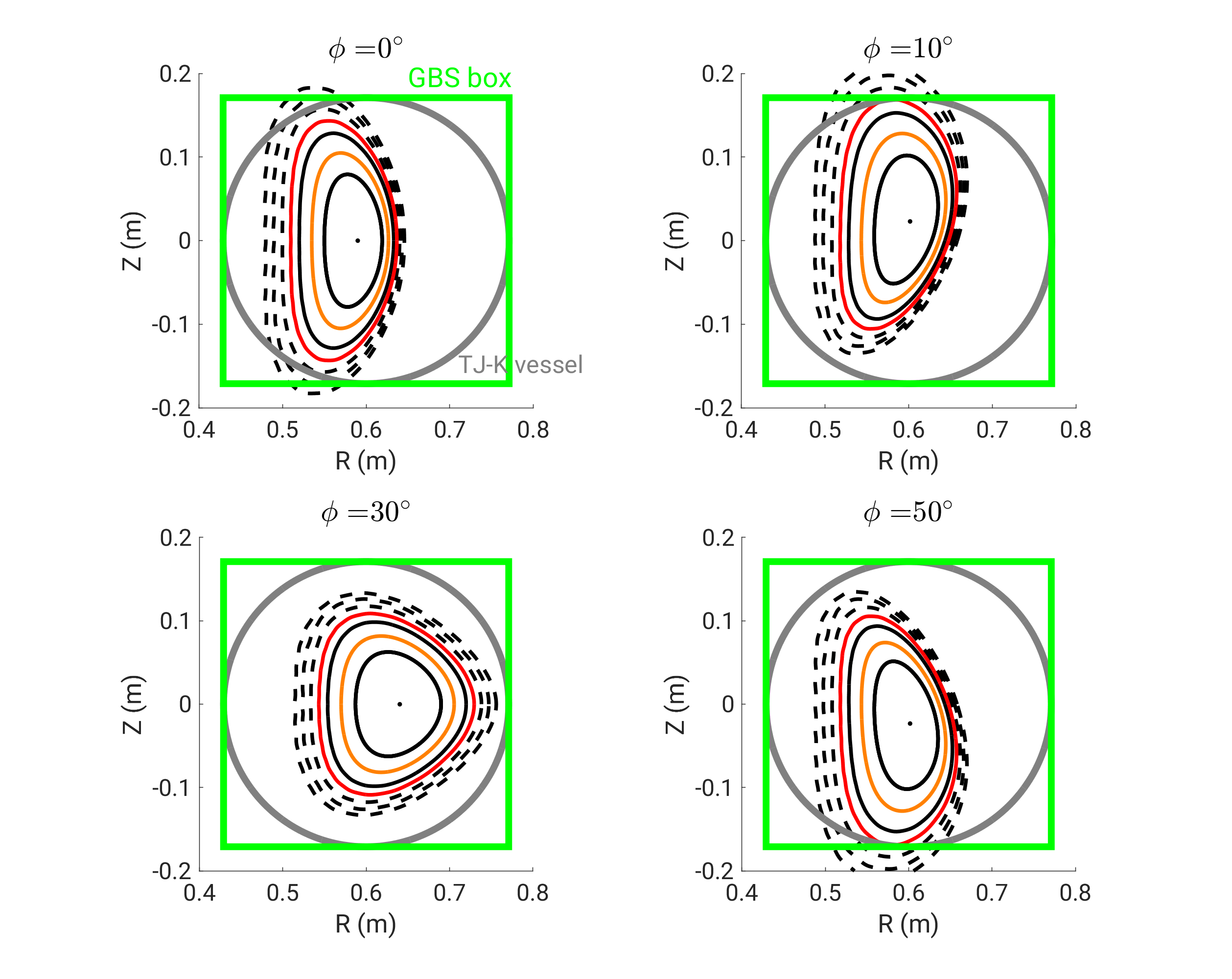}
 \caption{Poincar\'e plots of the TJ-K magnetic field at four different toroidal angles. The circular vessel is depicted in grey and the GBS simulation box in green. Closed flux surfaces are represented by continuous lines and open flux surfaces by dashed black lines. These are separated by the LCFS in red. The orange line refers to the surface where density and potential fluctuation measurements are performed (this is referred to as the \textit{reference surface}).}
 \label{fig:TJK_poincare}
\end{figure}

TJ-K is a six-field period stellarator with a major radius of 0.6\,m and a minor radius of, approximately, 0.1\,m~\cite{TJ-K}. The vacuum magnetic field is generated by a helical coil that loops around the vessel six times, and two vertical-field circular coils, as shown in Fig.~\ref{fig:TJK_esquema}. The magnetic field strength is, approximately, 70\,mT. The Poincar\'e plots at four different toroidal angles are shown in Fig.~\ref{fig:TJK_poincare} with the toroidal vessel depicted in grey. Continuous lines correspond to closed flux surfaces, while dashed lines correspond to open flux surfaces. The last closed flux surface (LCFS) is represented in red. In fact, the plasma is limited at the top and bottom part of the vessel at two different poloidal planes for every field period. The profile of the rotational transform is approximately flat, with a value $\iota=0.28$ at the magnetic axis (henceforth defined by the coordinates $R_{\text{axis}}$ and $Z_{\text{axis}}$, that vary along the toroidal angle $\phi$).  

In TJ-K, the plasma breakdown and heating is achieved with electron cyclotron resonant heating (ECRH), using a 2.45 GHz microwave system with up to 6 kW heating power~\cite{kohn_microwaves}. When working with hydrogen or helium gases, this yields typical line-averaged plasma densities of order $10^{17}\,\text{m}^{-3}$ and the electron temperature around $10\,\text{eV}$, while ions are cold (the ion temperature is less than 1\,eV). This results in a plasma collisionality $\nu^*\simeq10$, with $\nu^*$ defined as the ratio between the trapped particle collision frequency and the banana bounce frequency. As a consequence, TJ-K plasmas are in the Pfirsch-Schlüter regime~\cite{helander_sigmar_book}. 

The TJ-K Langmuir probes enable measurements with good spatial resolution. The radial profiles of density and electron temperature are measured with a radially movable Langmuir probe, and the radial profile of the electrostatic potential with an emissive probe. In addition, density and plasma potential fluctuations are measured with two multi-probe arrays consisting of 64 Langmuir probes each, located at two different toroidal locations, one at an outer port (OPA) at $\phi=30^{\circ}$ and the other at a top port (TPA) at $\phi=10^{\circ}$. The probe tips are aligned to the same flux surface as shown in Fig.~\ref{fig:TJK_esquema}. This surface, referred to as the \textit{reference surface} in the following, corresponds to the orange surface in the Poincar\'e plots of Fig.~\ref{fig:TJK_poincare}. These two arrays allow for detailed measurements of the potential and density fluctuations as a function of the poloidal angle, $\theta$. In particular, the radial $\text{E}\times\text{B}$ turbulent particle flux is evaluated as

\begin{equation}
    {\Gamma}_{E\times B}^{\text{exp}}(\theta_i) = \frac{1}{B_i}\left<-\frac{\widetilde{\Phi}_{\text{fl},i+1}-\widetilde{\Phi}_{\text{fl},i-1}}{2 \Delta y}\widetilde{I}_{\text{sat},i}\right>_t,
    \label{eq:flux_ExB_experiment}
\end{equation}
where $\widetilde{\Phi}_{\text{fl},i}$ and $\widetilde{I}_{\text{sat},i}$ are, respectively, the floating potential and ion saturation current fluctuations as measured by the $i$-th probe at the poloidal position $\theta_i$, $B_i$ is the magnetic field strength at the same position, and $\Delta y\approx8\text{ mm}$ is the distance between adjacent probe tips covering the flux surface. The 64 tips alternate between measurements of $\widetilde{\Phi}_{\text{fl}}$ and $\widetilde{I}_{\text{sat}}$, hence the tips $\{i-1,i,i+1\}$ are used to compute the flux at $\theta_i$. The temporal average is carried out over 1024\,ms of the data sampled at 1\,MHz, resulting in a very small uncertainty of the mean value. Systematic errors due to possible probe misalignments can be estimated to have maximum relative values of 13\%~\cite{TJK_PRL_2011}. The use of the floating potential instead of the plasma potential in the evaluation of ${\Gamma}_{E\times B}^{\text{exp}}$ is justified by the negligible temperature fluctuations present in the experiment~\cite{TJK_Te_fluctuations}.


\section{The GBS simulation}

GBS~\cite{paolo_GBS,GBS_cite_Maurizio,halpern} is a three-dimensional, global, two-fluid, flux-driven code that solves the drift-reduced Braginskii equations~\cite{zeiler}, valid in the high-collisionality regime that often characterizes the plasma boundary of magnetic fusion devices as well as the core of low-temperature devices such as TJ-K. GBS evolves all quantities in time, without separation between equilibrium and fluctuating parts. We consider here the cold ion and the electrostatic limits, we apply the Boussinesq approximation~\cite{paolo_GBS} and neglect gyroviscous terms as well as the coupling to the neutral dynamics~\cite{wersal_neutrals}. 
Within these approximations, the drift-reduced model evolved by GBS for the simulation considered in this paper is:

\begin{equation}
    \pt{n} = -\frac{\rorhos}{B}\left[\Phi,n\right] - \nablapar(n\vpare) + \frac{2}{B}\left[C(p_e)-nC(\Phi)\right] + D_n\nablaperp^2n + D_n^{\parallel}\nablapar^2n +  \mathcal{S}_n 
    \label{eq:density}
\end{equation}

\begin{equation}
\begin{aligned}
    \pt{T_e} = &-\frac{\rorhos}{B}\left[\Phi,T_e\right] - \vpare\nablapar T_e + \frac{4T_e}{3B}\left[\frac{C(p_e)}{n}+\frac{5}{2}C(T_e)-C(\Phi)\right]\\ &+ \frac{2T_e}{3n}\left[0.71\nablapar j_{\parallel}-n\nablapar\vpare\right] + D_{T_e}\nablaperp^2T_e+\chi_{\parallel e}\nablapar^2T_e + \mathcal{S}_{T_e}
\end{aligned}
\end{equation}

\begin{equation}
\begin{aligned}
    \pt{\vpare} = &-\frac{\rorhos}{B}\left[\Phi,\vpare\right]-\vpare\nablapar\vpare + \frac{m_i}{m_e}\left[\nu j_{\parallel}+\nablapar\Phi-\frac{\nablapar p_e}{n} - 0.71\nablapar T_e\right]\\ &+ \eta_{0e}\nablapar^2\vpare + D_{\vpare}\nablaperp^2\vpare
    \label{eq:vpare}
\end{aligned}
\end{equation}

\begin{equation}
    \pt{\vpari} = -\frac{\rorhos}{B}\left[\Phi,\vpari\right]-\vpari\nablapar\vpari - \frac{1}{n}\nablapar p_e +\eta_{0i}\nablapar^2\vpari + D_{\vpari}\nablaperp^2\vpari
\end{equation}

\begin{equation}
    \pt{\omega} = -\frac{\rorhos}{B}\left[\Phi,\omega\right] -\vpari\nablapar\omega + \frac{B^2}{n}\nablapar j_{\parallel} + \frac{2B}{n}C(p_e) + D_{\omega}\nablaperp^2\omega + D_{\omega}^{\parallel}\nablapar^2\omega
    \label{eq:vorticity}
\end{equation}

\begin{equation}
    \nablaperp^2\Phi = \omega
    \label{eq:potential}
\end{equation}
In Eqs.~(\ref{eq:density}-\ref{eq:potential}) all quantities are normalized to reference values. Density $n$ and electron temperature $T_e$ are normalized to the reference values $n_0$ and $T_{e0}$; electron parallel velocity $\vpare$ and ion parallel velocity $\vpari$ are both normalized to the sound speed $c_{s0}=\sqrt{T_{e0}/m_i}$; vorticity $\omega$ and the electrostatic potential $\Phi$ are normalized to $T_{e0}/(e\rhos^2)$ and $T_{e0}/e$; time is normalized to $R_0/c_{s0}$, where $R_0$ is the machine major radius; perpendicular and parallel lengths are normalized to the ion sound Larmor radius, $\rhos=\sqrt{T_{e0}m_i}/(eB_0)$, and $R_0$, respectively. The normalized parallel current is $j_{\parallel}=n(\vpari-\vpare)$ and the magnetic field $B$ is normalized to its norm at the magnetic axis, $B_0$.

The dimensionless parameters appearing in the equations are the normalized ion sound Larmor radius $\rho_*=\rhos/R_0$; the normalized electron and ion parallel diffusivities, $\chi_{\parallel e}$ and $\chi_{\parallel i}$ (here considered constant); the normalized electron and ion viscosities, $\eta_{0e}$ and $\eta_{0i}$, which we also set to constant values; and the normalized Spitzer resistivity $\nu=\nu_0T_e^{3/2}$ with $\nu_0$ given in Ref. \cite{maurizio_turbulent_regimes}. Small numerical diffusion terms such as $D_n\nablaperp^2n$ and $D_n^{\parallel}\nablapar^2n$ (and similar for the other fields) are introduced to improve the numerical stability of the code, and the simulation results show that they lead to significantly lower perpendicular transport than the turbulent one. The terms $\mathcal{S}_n$ and $\mathcal{S}_{T_e}$ denote the normalized sources of density and electron temperature. Magnetic presheath boundary conditions, described in Refs. \cite{joaquim_BCs,annamaria_BCs}, are applied to all quantities at the end of the field lines intersecting the walls, except for density and vorticity. These satisfy $\partial_sn=0$ and $\omega=0$, respectively, where $s$ is the direction normal to the wall. 

The normalized geometrical operators appearing in Eqs. (\ref{eq:density}-\ref{eq:potential}) are the parallel gradient $\nabla_{\parallel}u = \boldsymbol{b}\cdot\boldsymbol{\nabla}u$, the Poisson brackets $[\Phi,u]=\boldsymbol{b}\cdot\left[\boldsymbol{\nabla}\Phi\times\boldsymbol{\nabla} u\right]$, the curvature operator $C(u) = (B/2)\left[\boldsymbol{\nabla}\times(\boldsymbol{b}/B)\right]\cdot\boldsymbol{\nabla}u$, the parallel Laplacian $ \nabla_{\parallel}^2u = \boldsymbol{b}\cdot\boldsymbol{\nabla}(\boldsymbol{b}\cdot\boldsymbol{\nabla}u)$ and the perpendicular Laplacian $\nabla_{\bot}^2u = \boldsymbol{\nabla}\cdot\left[(\boldsymbol{b}\times\boldsymbol{\nabla}u)\times\boldsymbol{b}\right]$.

The simulation domain is a torus of radius $R_0$ with a rectangular cross-section. Since the vessel of the experiment has, instead, a circular cross-section, we choose the size of the domain in order to limit the plasma at the same positions as in the experiment. This simulation domain is shown in Fig.~\ref{fig:TJK_poincare} (green line). The physical model in Eqs.~(\ref{eq:density}-\ref{eq:potential}) is discretized using a regular cylindrical grid $(R,\phi,Z)$, with $R$ the radial coordinate, $\phi$ the toroidal angle and $Z$ the vertical coordinate. Equations (\ref{eq:density}-\ref{eq:vorticity}) are advanced in time with an explicit Runge-Kutta fourth-order scheme, while spatial derivatives are computed with a fourth-order finite difference scheme.

The magnetic field was initially computed numerically using the MAKEGRID code from the STELLOPT package~\cite{STELLOPT}, which uses the Biot-Savart law to determine the magnetic field at a specified location, using the coil geometry and coil currents as an input. However, since the vessel of the experiment, as well as the helical coil, fall inside the GBS domain, singularities in the magnetic field at the position of the coil appear, hindering an appropriate implementation of the geometrical operators. To circumvent this issue, we select one of the open flux surfaces (obtained with the FIELDLINES code from the same package) and provide it as an input to REGCOIL~\cite{regcoil}. This code seeks a surface current distribution on an arbitrary toroidal surface, which we choose to enclose the GBS rectangular domain, such that $\mathbf{B}\cdot\mathbf{n}=0$ at the provided surface. Furthermore, since $\nabla\times\mathbf{B}=0$ in vacuum, the problem is reduced to a Laplace equation with Neumann boundary conditions, that admits a unique solution (up to a re-scaling factor). This results in a magnetic field obtained by REGCOIL that is exactly the same as that of TJ-K inside the chosen surface (apart from the re-scaling factor). With this approach, we obtain a non-singular vacuum magnetic field that coincides with that of TJ-K from the magnetic axis up to the selected open flux surface. 

\begin{figure}[]
 \centering
 \includegraphics[width=8cm]{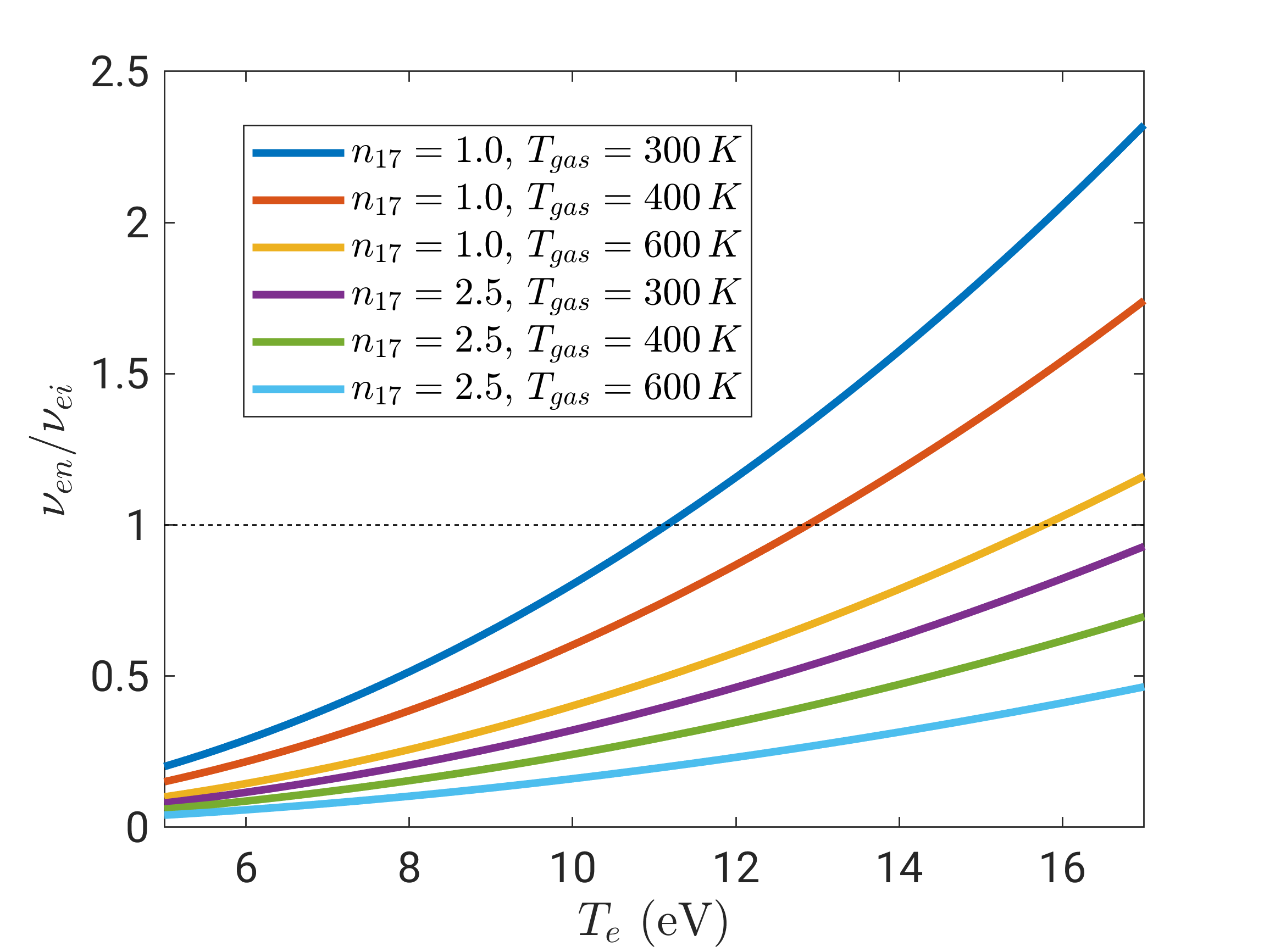}
 \caption{Ratio between electron-neutral and electron-ion collision frequencies at different plasma densities and gas temperatures, $T_{gas}$. We indicated with $n_{17}$ the plasma density in units of $10^{17}\,\text{m}^{-3}$. The neutral pressure was assumed to be $3\,\text{mPa}$, the gas pressure of the TJ-K discharges presented in this paper.}
 \label{fig:TJK_collision_times}
\end{figure}

TJ-K plasmas are not fully ionized. Hence, the electron-neutral collisions could affect the drift-reduced Braginskii model, in particular the parallel friction and parallel thermal conduction terms. This can be assessed by quantifying the ratio $\nu_{\text{en}}/\nu_{\text{ei}}$, where $\nu_{ei}$ is the Coulomb electron-ion collision frequency~\cite{braginskii} and $\nu_{en}=n_n\sigma_{\text{en}} v_{Te}$ is the electron-neutral collision frequency, with $n_n$ the neutrals density, $\sigma_{\text{en}}$ the momentum transfer cross-section for electrons impacting neutrals and $v_{Te}$ the electron thermal velocity. The ratio is shown in Fig.~\ref{fig:TJK_collision_times} for electron temperatures ranging between 5\,eV and 17\,eV, for which $\sigma_{\text{en}}\approx10^{-19}\text{m}^{-2}$~\cite{cross_section_helium}. The different curves refer to different plasma densities and neutral temperatures, and show that the electron-neutrals interaction can become important at large electron temperatures. Nevertheless, in this work we assume that these interactions do not play a role.

We consider that the source of density is due to ionization processes. This is modeled as $\mathcal{S}_n = n_n n\left<\sigma_{\text{ion}} v\right>_v$. However, because the neutral pressure in TJ-K is approximately constant throughout the plasma volume and ionization is due to fast electrons (whose density is assumed proportional to the plasma density), the density source is recast as $\mathcal{S}_n=\alpha_n n$, with $\alpha_n$ a constant. We neglect recombination processes because of the typical TJ-K temperatures.

\begin{figure}[]
 \centering
 \includegraphics[width=16cm]{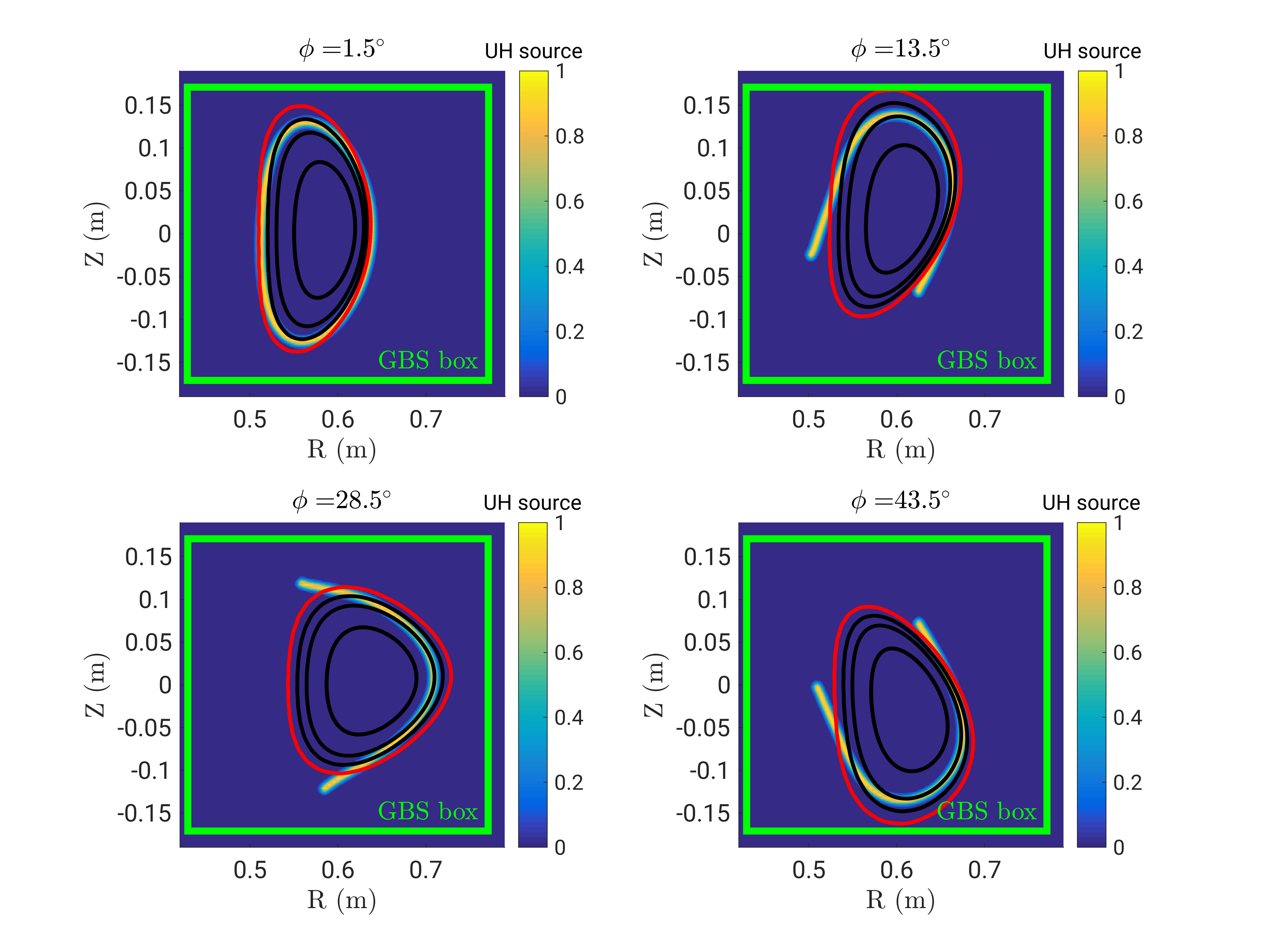}
 \caption{The UH resonant layer, $P(\mathbf{r})$, considered in the simulation. The layer is obtained by matching the UH frequency with the frequency of the antenna, $\sqrt{\omega_{\text{pe}}^2+\omega_{\text{ce}}^2} = 2\pi f_{\text{antenna}}$, being $f_{\text{antenna}}=2.45\text{ GHz}$.}
 \label{fig:UH_source}
\end{figure}

The electron temperature source is composed of the external energy input power and a term associated with the particle source, leading to

\begin{equation}
    \mathcal{S}_{Te}=\frac{R_0/c_{s0}}{T_{e0}}\left[\frac{2}{3}\frac{\mathcal{P}}{n_0n} - \frac{2}{3}n_0\left<\sigma v\right>_{\text{ion}}\left(E_{\text{ion}}+\frac{3}{2}T_e\right)\right],
    \label{eq:temp_source_1}
\end{equation}
where $\mathcal{P}$ is the ECRH input power density, $\left<\sigma v\right>_{\text{ion}}$ is the ionization reaction-rate, which is of the order of $10^{-15}\,\text{m}^3/\text{s}$ at $10\text{ eV}$~\cite{lechte_ionization_coeffs}, and $E_{\text{ion}}$ is the ionization energy. As described in Ref.~\cite{kohn_microwaves}, power deposition in TJ-K occurs at the upper hybrid (UH) resonance layer. The space-dependent input power $\mathcal{P}$ is given by

\begin{equation}
   \mathcal{P} = \frac{\mathcal{P}_{\text{ant}}}{\int_VP(\mathbf{r})d\mathbf{r}}P(\mathbf{r}),
\end{equation}
where $\mathcal{P}_{\text{ant}}=1.8 \text{ kW}$ is the power launched by the antennas in the TJ-K discharges considered here and $P(\mathbf{r})$ is the UH resonant layer, shown in Fig.~\ref{fig:UH_source} at different poloidal planes. $P(\mathbf{r})$ is obtained by matching the UH frequency with the antenna frequency, $\sqrt{\omega_{\text{pe}}^2+\omega_{\text{ce}}^2} = 2\pi f_{\text{antenna}}$, and then assuming a Gaussian deposition profile located around the isocountours that solve this relation. The temperature source in Eq.~(\ref{eq:temp_source_1}) can thus be written as

\begin{equation}
    \mathcal{S}_{Te}=\alpha_{Te}\left[\frac{P(\mathbf{r})}{n}-3\times10^{-4}E_{\text{ion}}^{\text{eV}}(1+T_e)\right].
    \label{eq:source_tempe}
\end{equation}
For helium, we consider the first ionization energy, $E_{\text{ion}}^{\text{eV}}=24.6\,\text{eV}$. The constants $\alpha_n$ and $\alpha_{Te}$ are adjusted such that the peak values of density and temperature are close to the reference values, $n\approx 1$ and $T_e\approx1$. This adjustment accounts for the uncertainties on the averaging of the ionization reaction-rate and on the effective absorption by the plasma of the ECRH power.

Turning now to the simulation parameters, we consider $\rorhos=60$ and $\nu_0=5\times10^{-4}$ by using the reference values $T_{e0}=10\text{ eV}$ and $n_0=10^{17}\text{m}^{-3}$. We use an atomic number $Z=1$ since helium is mostly single-ionized at the reference temperature. We further use $m_i/m_e=900$, $\chi_{\parallel e}=0.5$, $\eta_{0 e,i}=1.0$, $D_n=D_{Te}=D_{\vpare}=D_{\vpari}=D_{\omega}=0.2$, $D_{n}^{\parallel}=0.18$, $D_{\omega}^{\parallel}=0.01$, $\alpha_n=0.03$ and $\alpha_{Te}=0.7$. Concerning the numerical parameters, the simulation is performed with a time-step of $2.0\times 10^{-5}R_0/c_{s0}$ and a grid resolution of $\Delta R=\Delta Z=0.5\rho_{s0}$ and $\Delta\phi=2\pi/(20\times6)$, i.e., 20 poloidal planes per field period. A convergence test in the toroidal direction, made by increasing the number of planes from 20 to 30 per field period, show results similar to the one presented here. 

The simulation of TJ-K we consider is started from an initial state with background noise and, after a transient, reaches a quasi-steady state where sources, parallel and perpendicular transport, and losses at the vessel balance each other. The analysis of the simulations results is performed during this quasi-steady state.

\section{Simulation results and comparison with the experiment}

\begin{figure}[]
 \centering
 \includegraphics[width=16cm]{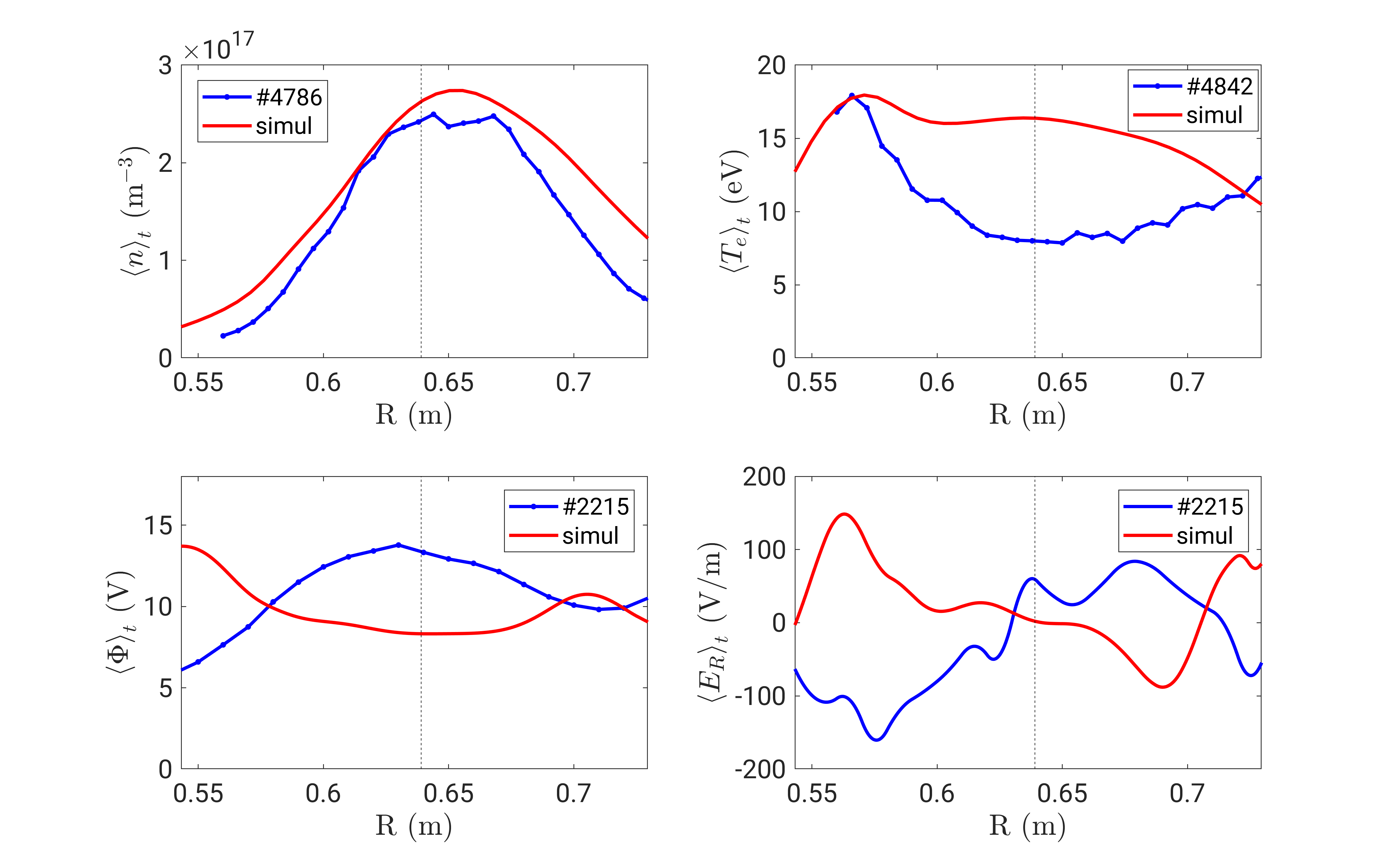}
 \caption{Radial profiles of the time-averaged (equilibrium) density, electron temperature, potential and radial electric field. Profiles are at $Z=0$ and $\phi=30^{\circ}$ (OPA). The vertical dashed line indicates the position of the magnetic axis, $R_{\text{axis}}$.}
 \label{fig:equilibrium_profiles}
\end{figure}

We start by comparing the one-dimensional time-averaged (equilibrium) profiles of density, temperature, potential and radial electric field, $E_R=-\nabla\Phi\cdot\mathbf{\widehat{e}_R}$, along the $R$ direction, at $Z=0$ and $\phi=30^{\circ}$. The comparison is shown in Fig.~\ref{fig:equilibrium_profiles}. The peak value as well as the profile of the simulated density agree well with the experiment. The experimental temperature profile is hollow and this is attributed to the fact that the resonance layer of the UH is located at the edge (see Fig.~\ref{fig:UH_source}). On the other hand, the simulation displays a temperature profile that decays for $R>R_{\text{axis}}$. Furthermore, the temperature on axis is larger in the simulation than in the experiment. We note that the simulated temperature profile is sensitive to $\alpha_{Te}$. Reducing this parameter lowers the value of the temperature, but the equilibrium gradients change, eventually suppressing turbulence. The magnitude of the electrostatic plasma potential is similar in both experiment and simulation (between 8 and 14\,V), but the radial dependencies are different. In fact, the simulation reveals a hollow potential profile, whereas in the experiment the profile increases as the axis is approached. This makes the profiles of $E_R$ also different, even if its order of magnitude is correctly captured.

\begin{figure}[]
 \centering
 \includegraphics[width=16cm]{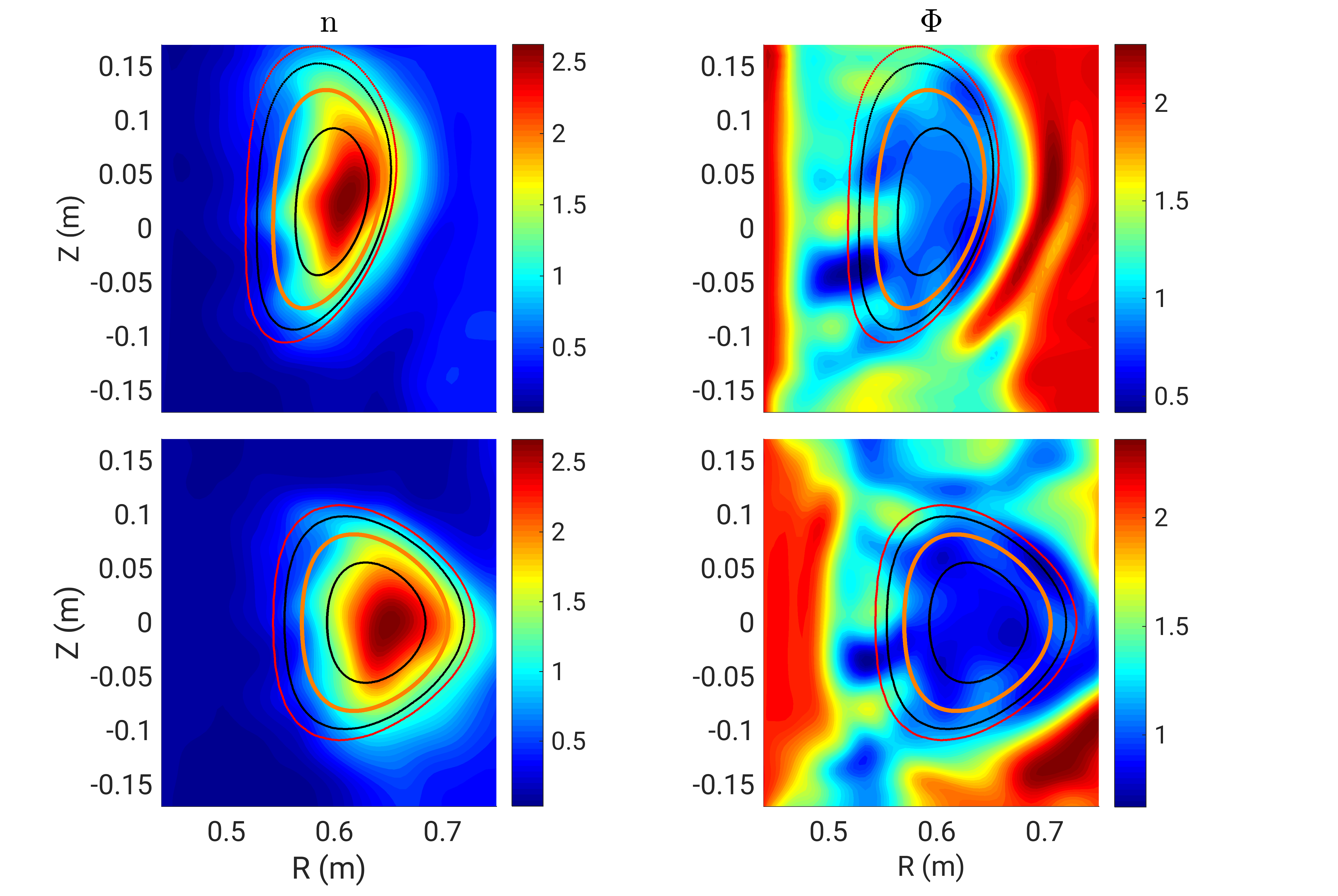}
 \caption{Snapshot of density, $n$ (left) and electrostatic potential, $\Phi$ (right), in the quasi-steady state of the GBS simulation. Top and bottom correspond to the toroidal planes $\phi=10^{\circ}$ (TPA) and $\phi=30^{\circ}$ (OPA), respectively. The reference surface is indicated with an orange line.}
 \label{fig:snapshots_n_phi}
\end{figure}

\begin{figure}[]
 \centering
 \includegraphics[width=20cm]{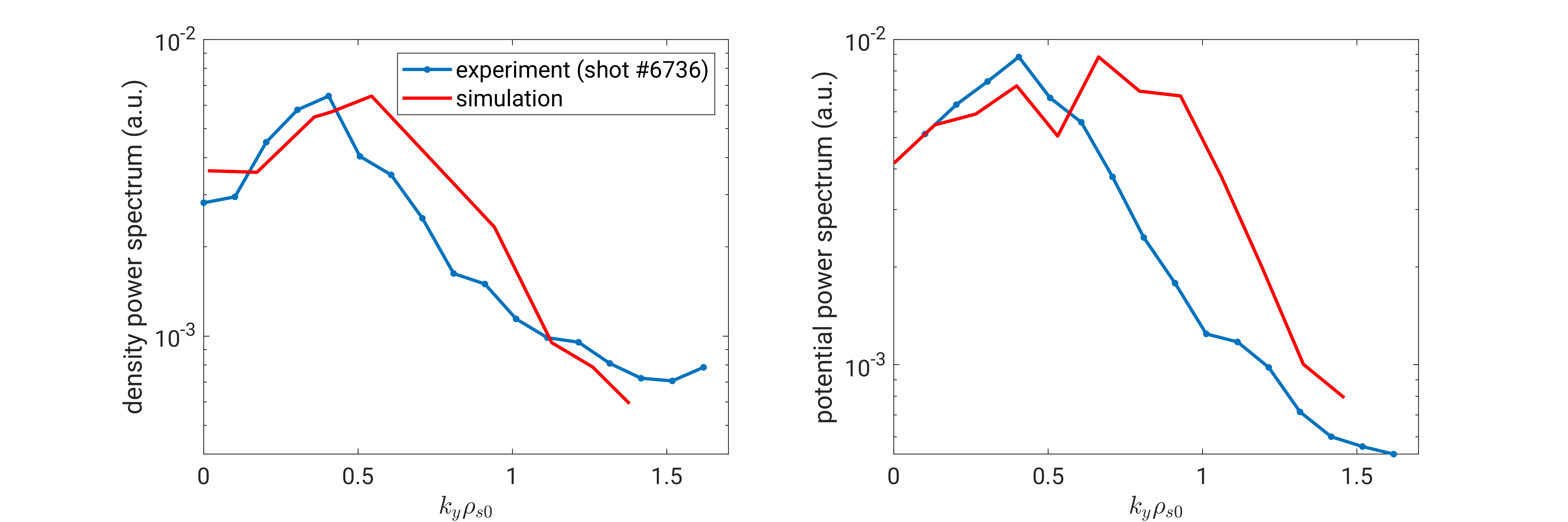}
 \caption{Power spectrum of density (left) and electrostatic potential (right) of the fluctuations. The spectra are computed by Fourier transforming density fluctuations along $y$ at OPA ($\phi=30^{\circ}$). The simulation spectrum is normalized such that the maxima of experiment and simulation are the same.} 
 \label{fig:power_spectra}
\end{figure}

Snapshots of the density and electrostatic potential on two different poloidal planes, $\phi=10^{\circ}$ and $\phi=30^{\circ}$, are shown in Fig.~\ref{fig:snapshots_n_phi}. We observe that a low-$m$ coherent mode dominates the plasma dynamics, where $m$ is the poloidal mode number. We Fourier decompose the fluctuations along the $y$-coordinate, where $y$ is the arc length along the poloidal projection of the reference surface. Fig.~\ref{fig:power_spectra} (left) shows the power spectrum of density fluctuations obtained at $\phi=30^{\circ}$ and compares it with the experiment. The simulation retrieves the coherent mode present in the experiment at $k_y\rho_{s0}\approx0.4$, which corresponds to an $m=4$ structure. The experimental spectrum decays with a power law $(k_y\rho_{s0})^{-1.9}$, consistent with the inverse cascade in two-dimensional fluid turbulence~\cite{TJK_energy_cascade}. On the other hand, the observed power law in the simulation, $(k_y\rho_{s0})^{-1.2}$, is slightly underestimated with respect to the experiment. The power spectrum at $\phi=10^{\circ}$ (not shown) is similar to the one at $\phi=30^{\circ}$, both in the experiment and simulation, an expected feature since turbulence is field-aligned. 
Regarding the power spectrum of the electrostatic potential (right panel in Fig.~\ref{fig:power_spectra}), the simulation has two dominant coherent modes at $k_y\rho_{s0}\approx0.4$ and $k_y\rho_{s0}\approx0.7$, although only the first one is present in the experiment. In addition, the spectrum shows a slightly faster decay in the simulation than in the experiment. 

\begin{figure}[]
 \centering
 \includegraphics[width=16cm]{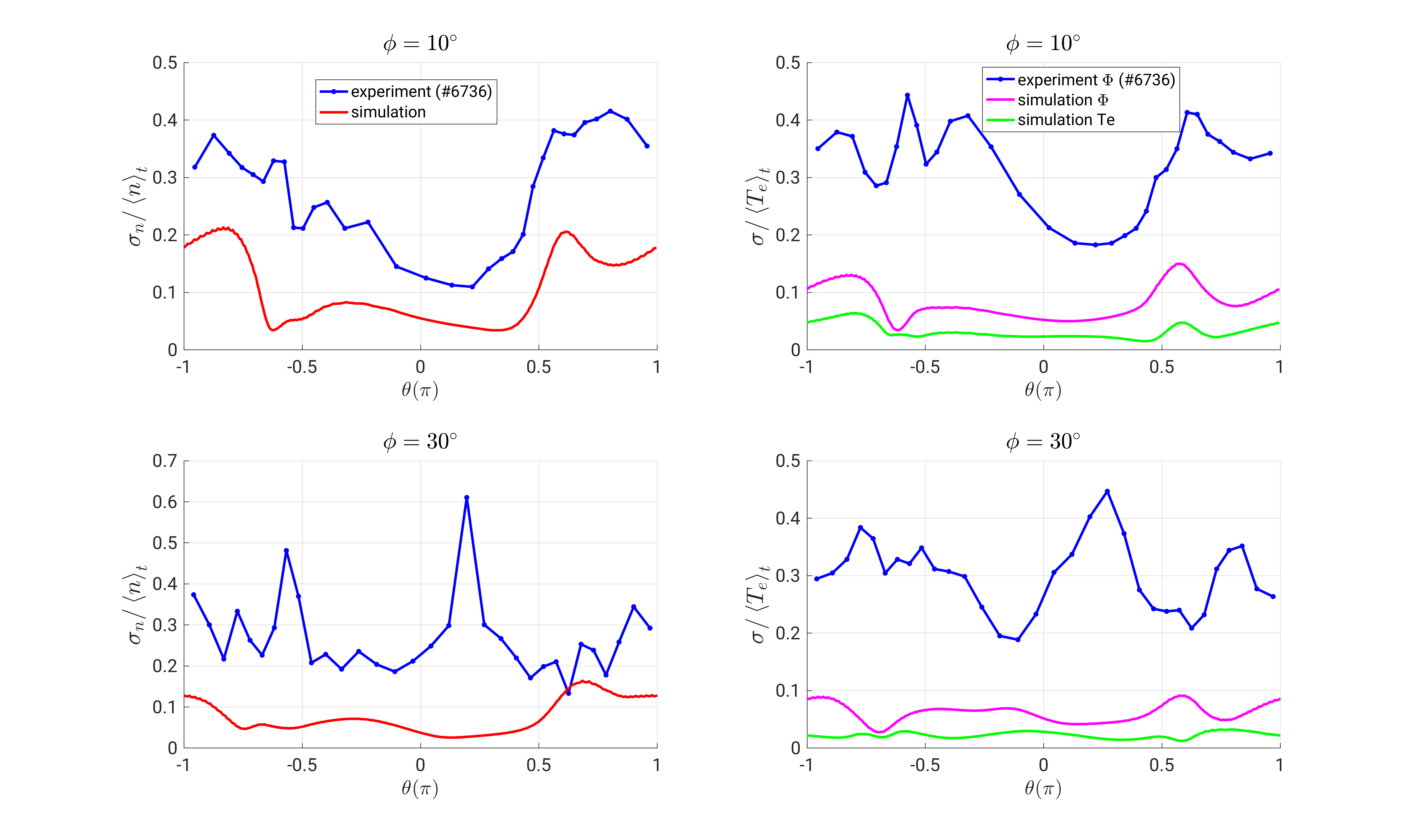}
 \caption{Density (left) and potential and electron temperature (right) fluctuation levels as a function of the poloidal angle, $\theta$, at $\phi=10^{\circ}$ (top) and $\phi=30^{\circ}$ (bottom).}
 \label{fig:fluctuation_level_comparison}
\end{figure}

In Fig.~\ref{fig:fluctuation_level_comparison}, we show the density and potential fluctuation levels as a function of the poloidal angle on the reference surface, for the two considered poloidal planes. The fluctuation level of the density is calculated as the standard deviation, $\sigma_n$, normalized to the radially dependent equilibrium value, $\left<n\right>_t$. The normalizing factor of the potential is the equilibrium electron temperature, $\left<T_e\right>_t$, to avoid possible singularities. The simulation and experiment display a similar poloidal dependence at $\phi=10^{\circ}$. However, the simulation underestimates the fluctuation levels by a factor two or more. This is similar to previous validation studies with GBS and other fluid codes, where a lower level of fluctuations with respect to experiments is reported~\cite{torpex_galassi,diego_X21}. 
At $\phi=30^{\circ}$, the simulated fluctuations are also significantly smaller than in the experiment and, in addition, fluctuations do not peak at the same poloidal angle.
Regarding electron temperature fluctuations, they are negligible in TJ-K~\cite{TJK_Te_fluctuations}. In the simulation, although not being negligible, they are smaller than density and potential fluctuations, as demonstrated in Fig.~\ref{fig:fluctuation_level_comparison}.

\begin{figure}[]
 \centering
 \includegraphics[width=16cm]{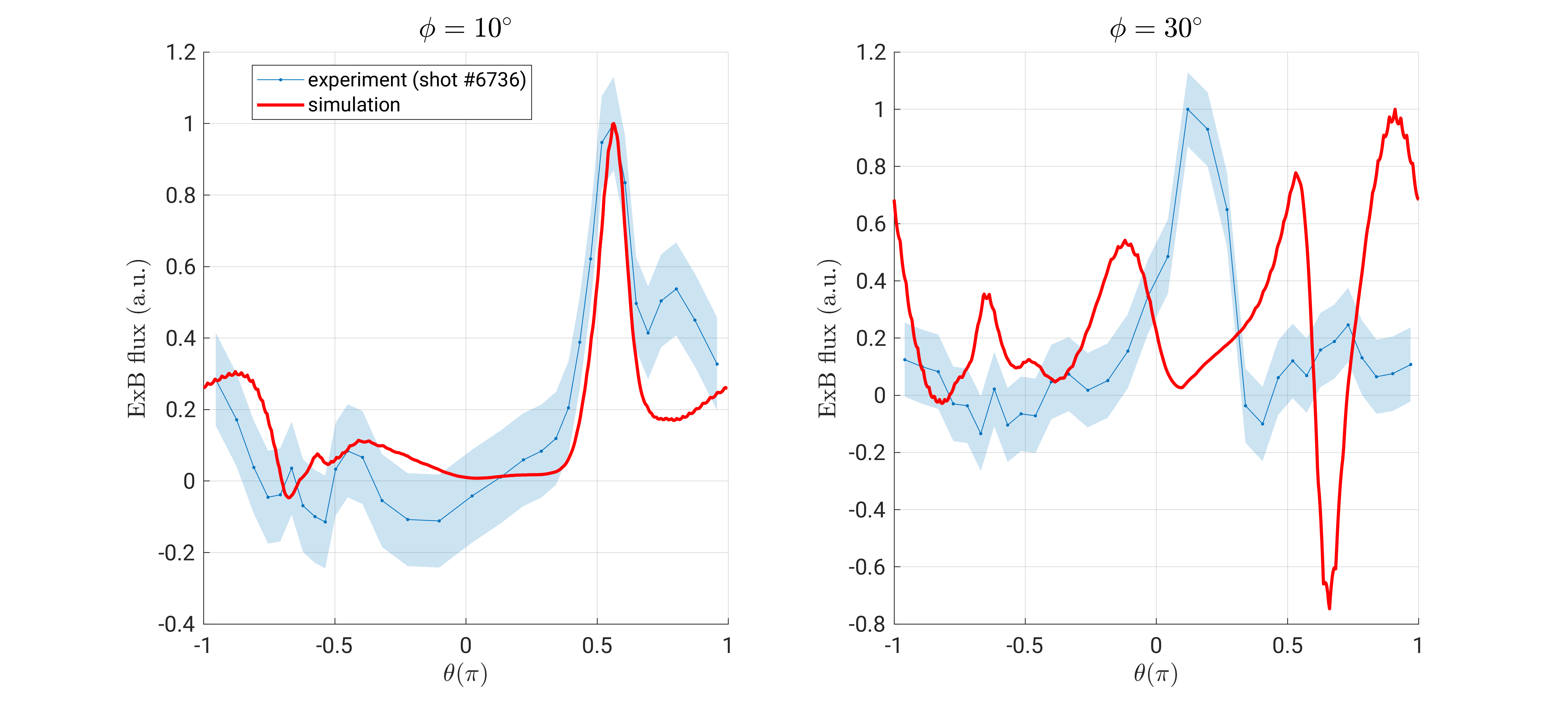}
 \caption{Comparison of the $\text{E}\times \text{B}$ flux as given by experiment and simulation at $\phi=10^{\circ}$ (TPA) and $\phi=30^{\circ}$ (OPA).}
 \label{fig:fluxes_comparison}
\end{figure}

The simulated radial turbulent $\text{E}\times\text{B}$ particle flux,

\begin{equation}
    {\Gamma}_{E\times B}=\left<\widetilde{n}\widetilde{V}^r_{E\times B}\right>_t=-\left<\frac{\widetilde{n}}{B}\left(\nabla\widetilde{\Phi}\times\mathbf{b}\right)_r\right>_t,
    \label{eq:ExB_flux}
\end{equation}
where $r$ denotes the direction normal to the flux surface, is compared with the experiment. The comparison is shown in Fig.~\ref{fig:fluxes_comparison}. Both fluxes are normalized to their peak values since the experimental value of transport is based on the ion saturation current. The simulation and experimental results at $\phi=10^{\circ}$ agree considerably better than at $\phi=30^{\circ}$. At $\phi=10^{\circ}$ the simulation retrieves the transport peak occurring at around $\theta=\pi/2$, while the peak at $\theta=0.2\pi$ is not retrieved at $\phi=30^{\circ}$. This is partially explained by the different peak position of the fluctuation levels observed at $\phi=30^{\circ}$ (see Fig.~\ref{fig:fluctuation_level_comparison}). In addition, the cross-phase has an important role in setting the level of transport. In fact, the simulation shows a peak of the flux with a negative value, revealing that the phase-difference between density and potential also varies poloidally, hence having an important role in the turbulent transport asymmetries. It is worth mentioning that in the experiment there is a good correlation between the position where turbulent transport is maximum and the region where normal and geodesic curvatures are, respectively, negative and positive~\cite{TJK_PRL_2011}, something that could not be verified with the simulations.

\begin{figure}[]
 \centering
 \includegraphics[width=16cm]{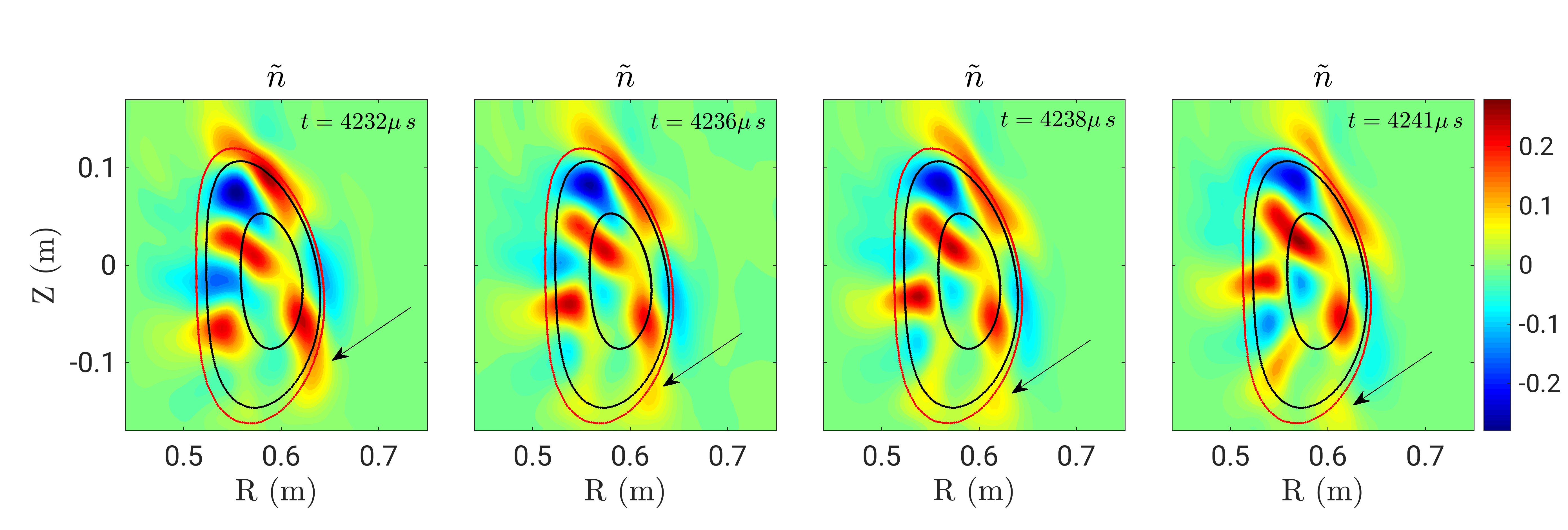}
 \caption{Time sequence of density fluctuations in the simulation. The arrow points to a region where a large coherent structure, originated inside the LCFS, detaches and propagates towards the SOL within a few $\mu s$.}
 \label{fig:blob_sequence}
\end{figure}

Coherent structures originating from inside the LCFS and propagating outwards into the scrape-off layer (SOL) are typically observed in TJ-K by means of a 2D movable probe and/or a fast camera~\cite{TJK_PRL_2009,TJK_blobs_2013,TJK_blobs_2016}. The origin of these structures is attributed to drift-waves turbulence. 
The presence of coherent structures propagating from the closed field line region towards the SOL is visible also in the simulation. This is shown in Fig.~\ref{fig:blob_sequence}, where a time sequence of the density fluctuations is shown. The formation of this coherent structure occurs in the region marked with an arrow within a few $\mu s$, as in the experiments~\cite{TJK_PRL_2009,TJK_blobs_2013}.

\section{Discussion and conclusions}

GBS captures the essential turbulence properties of TJ-K, for instance the $k_y$ spectra, as shown in Fig.~\ref{fig:power_spectra}. We note that both the spectrum and the fluctuation levels are robust to the source localization and strength. In fact, a simulation was carried out with density and temperature \textit{ad-hoc} sources of equal magnitude and localized around a closed flux surface in the proximity of the LCFS, showing very similar spectrum and fluctuation levels to the ones presented here. In addition, although not presented in this paper, a simulation with an hydrogen plasma also shows good agreement with the experimental $k_y$ spectrum. 

The equilibrium profiles of temperature, potential and electric field, which reveal significant differences between simulations and experiments, depend significantly on the details of the sources, whose localization and strength are, to some extent, uncertain. For example, using the \textit{ad-hoc} sources lead to an order magnitude difference of the $E_R$ values between simulation and experiment.
The origin of the electric field is unclear, both in the experiment and in the simulation. Since the ion temperature is small, neoclassical effects do not play a role in setting the electric field, in contrast to the case of high temperature stellarator plasmas~\cite{simakov_helander}. Moreover, a test where finite ion temperature effects are introduced in the system by setting $T_{i0}/T_{e0}=0.1$ as in the experiment, shows similar results to the cold ion simulation. 
We expect that an improvement of the source model could yield a better agreement between simulation and experimental radial electric field. In fact, plasma breakdown in TJ-K occurs at the electron cyclotron resonance layer~\cite{TJK_plasma_breakdown_EC}, and therefore the fast electrons produced at this resonance are responsible for the further ionization of the background gas. A proper modelling of this species could thus improve our comparison. Furthermore, as shown in Fig~\ref{fig:TJK_collision_times}, electron-neutrals interaction can become important at temperatures larger than 10\,eV. Since TJ-K temperatures are between 8\,eV and 17\,eV (see Fig.~\ref{fig:equilibrium_profiles}), it might be important to take into account such interactions in future investigations.  

Finally, we note that the role of boundary conditions is expected to be important in simulation of such small experiments such as TJ-K. The use of magnetic pre-sheath boundary conditions except for density and vorticity at the interface between the LCFS and the wall could affect the plasma dynamics and, for example, the peaking positions of the fluctuation levels.

To conclude, in this paper we present the first validation of the GBS code with a stellarator experiment. Overall, GBS captures the main features of turbulence, in particular the $k_y$ spectrum. The fluctuation levels are underestimated but their dependence on the poloidal angle is partially retrieved. Regarding the equilibrium quantities, GBS can retrieve the correct magnitudes of the potential and electric field, whose profiles are sensitive to the strength and localization of the sources. The equilibrium density profile agrees well with the experiment, but in the case of the electron temperature the value on axis is underestimated. As experimentally observed in TJ-K and confirmed by the GBS simulations presented here, the turbulent transport is mainly due to a low-$m$ mode, something that was also recently observed in a GBS simulation of a stellarator with an island divertor~\cite{gbs_stellarators}. This contrasts with the typical plasma turbulence in tokamaks, where more broad-band turbulence is observed. Ultimatelly, the validation of the presence of these low-$m$ modes calls for the detailed study of the difference between stellarators and tokamaks. 

\ack

The authors thank Caoxiang Zhu and Matt Landreman for all the help with REGCOIL. The simulations presented herein were carried out in part at the Swiss National Supercomputing Centre (CSCS) under the projects ID s1118 and s1182, and in part on the CINECA Marconi super computer. This work has been carried out within the framework of the EUROfusion Consortium, via the Euratom Research and Training Programme (Grant Agreement No 101052200 — EUROfusion) and funded by the Swiss State Secretariat for Education, Research and Innovation (SERI). Views and opinions expressed are however those of the author(s) only and do not necessarily reflect those of the European Union, the European Commission, or SERI. Neither the European Union nor the European Commission nor SERI can be held responsible for them.

\section*{References}
\bibliographystyle{unsrt}
\bibliography{bibliography}

\end{spacing}

\end{document}